\documentclass{emulateapj}
\usepackage{lscape}
\shorttitle{Spectroscopic Survey of GOODS-North}
\shortauthors{Reddy et~al.}
\slugcomment{Received 2006 June 23; Accepted 2006 September 01}

\begin{document}
\newcommand{\ebmv}{E(B-V)}
\newcommand{\sfr}{{\rm M}_{\odot} ~ {\rm yr}^{-1}}
\newcommand{\zmk}{(z-K)_{\rm AB}}
\newcommand{\jmk}{J-K_{\rm s}}
\newcommand{\rmk}{{\cal R}-\ks}
\newcommand{\gmr}{G-{\cal R}}
\newcommand{\umg}{U_{\rm n}-G}
\newcommand{\rmj}{{\cal R}-J}
\newcommand{\ugr}{U_{\rm n}G{\cal R}}
\newcommand{\rs}{{\cal R}}
\newcommand{\bzk}{BzK}
\newcommand{\kab}{K_{\rm AB}}
\newcommand{\ks}{K_{\rm s}}
\newcommand{\lya}{Lyman~$\alpha$}
\newcommand{\lyb}{Lyman~$\beta$}
\newcommand{\za}{$z_{\rm abs}$}
\newcommand{\ze}{$z_{\rm em}$}
\newcommand{\cmtwo}{cm$^{-2}$}
\newcommand{\nhi}{$N$(H$^0$)}
\newcommand{\degpoint}{\mbox{$^\circ\mskip-7.0mu.\,$}}
\newcommand{\kms}{\,km~s$^{-1}$}      % note leading thinspace
\newcommand{\minpoint}{\mbox{$'\mskip-4.7mu.\mskip0.8mu$}}
\newcommand{\peryr}{\mbox{$\>\rm yr^{-1}$}}
\newcommand{\secpoint}{\mbox{$''\mskip-7.6mu.\,$}}
\newcommand{\sqdeg}{\mbox{${\rm deg}^2$}}
\newcommand{\squig}{\sim\!\!}
\newcommand{\subsun}{\mbox{$_{\twelvesy\odot}$}}
\newcommand{\et}{{\rm et al.}~}

%\twocolumn[
\def\ltsima{$\; \buildrel < \over \sim \;$}
\def\simlt{\lower.5ex\hbox{\ltsima}}
\def\gtsima{$\; \buildrel > \over \sim \;$}
\def\simgt{\lower.5ex\hbox{\gtsima}}
\def\arcs{$''~$}
\def\arcm{$'~$}
\def\erf{\mathop{\rm erf}}
\def\erfc{\mathop{\rm erfc}}
\title{A SPECTROSCOPIC SURVEY OF REDSHIFT $1.4\la Z\la 3.0$ 
GALAXIES IN THE GOODS-NORTH FIELD: SURVEY DESCRIPTION, CATALOGS, AND
PROPERTIES\altaffilmark{1}} 
\author{\sc Naveen A. Reddy\altaffilmark{2,3}, Charles C. Steidel\altaffilmark{2}, Dawn K. Erb\altaffilmark{4}, Alice E. Shapley\altaffilmark{5}, and Max Pettini\altaffilmark{6}}

\altaffiltext{1}{Based on data obtained at the W.M. Keck
Observatory, which is operated as a scientific partnership among the
California Institute of Technology, the University of California, and
NASA, and was made possible by the generous financial support of the
W.M. Keck Foundation.}
\altaffiltext{2}{California Institute of Technology, MS 105--24, Pasadena, CA 91125}
\altaffiltext{3}{NOAO, 950 N Cherry Ave, Tucson, AZ 85719}
\altaffiltext{4}{Harvard-Smithsonian Center for Astrophysics, 60 Garden Street, Cambridge, MA 02138}
\altaffiltext{5}{Department of Astrophysical Sciences, Peyton Hall-Ivy Lane, Princeton, NJ 08544}
\altaffiltext{6}{Institute of Astronomy, Madingley Road, Cambridge CB3 OHA, UK}

%\slugcomment{DRAFT: \today}

\begin{abstract}

We present the results of a spectroscopic survey with the Low
Resolution Imaging Spectrograph on the Keck~I telescope of more than
$280$ star-forming galaxies and AGN at redshifts $1.4\la z\la 3.0$ in
the GOODS-North field.  Candidates are selected by their $\ugr$ colors
using the ``BM'' and ``BX'' criteria to target redshift $1.4\la z\la
2.5$ galaxies and the Lyman break criteria to target redshift $z\sim
3$ galaxies; combined these samples account for $\sim 25-30\%$ of the
{\cal R} and $\ks$-band counts to ${\cal R}=25.5$ and $\ks{\rm
(AB)}=24.4$, respectively.  The sample of $212$ BM/BX galaxies and
$74$ LBGs is presently the largest spectroscopic sample of galaxies at
$z>1.4$ in GOODS-N, and includes $19$ spectroscopically-confirmed
distant red galaxies (DRGs) with $\jmk>2.3$ (Vega).  Extensive
multi-wavelength data, including our very deep ground-based near-IR
imaging to $\ks{\rm (AB)}=24.4$, allow us to investigate the stellar
populations, stellar masses, bolometric luminosities ($L_{\rm bol}$),
and extinction of $z\sim 2$ galaxies.  Deep {\it Chandra} X-ray and
{\it Spitzer} IRAC and MIPS data indicate that the sample includes
galaxies with a wide range in $L_{\rm bol}$, from $\simeq
10^{10}$~L$_{\odot}$ to $>10^{12}$~L$_{\odot}$, and covering 4 orders of
magnitude in dust obscuration ($L_{\rm bol}/L_{\rm UV}$).  The sample
includes galaxies with a large dynamic range in evolutionary state,
from very young galaxies (ages $\simeq 50$~Myr) with small stellar
masses ($M^{\ast}\simeq 10^{9}$~M$_{\odot}$) to evolved galaxies (ages
$>2$~Gyr) with stellar masses comparable to the most massive galaxies
at these redshifts ($M^{\ast}>10^{11}$~M$_{\odot}$).  {\it Spitzer}
data indicate that the optical sample includes some fraction of the
obscured AGN population at high redshifts: at least 3 of 11 AGN in the
$z>1.4$ sample are undetected in the deep X-ray data but exhibit
power-law SEDs longward of $\sim 2$~$\mu$m (rest-frame) indicative of
obscured AGN.  The results of our survey indicate that rest-frame UV
selection and spectroscopy presently constitute the most time-wise
{\it efficient} method of culling large samples of high redshift
galaxies with a wide range in intrinsic properties, and the data
presented here will add significantly to the multi-wavelength legacy
of the GOODS survey.

\end{abstract}

%]

\section{Introduction}
\label{sec:intro}

Rapid advances in our understanding of galaxy evolution have been prompted
by the recognition that observations covering the full spectrum are
necessary to adequately interpret the physical nature of galaxies.
Multi-color {\it Hubble Space Telescope} ({\it HST}) imaging of two
otherwise inconspicuous fields in the high Galactic latitude sky
\citep{williams96, williams00} marked the inception of a decade dominated by
large-scale multi-wavelength surveys.  The two {\it Hubble} Deep
Fields are now encompassed or supplanted by other areas of the sky which are
the focus of a number of space and ground-based observations both
within and peripheral to the {\it Great Observatories Origins Deep
  Survey} (GOODS; \citealt{dickinsongoods03}).  Included among these
data are the deepest {\it Chandra} X-ray observations
\citep{alexander03}, {\it HST} ACS optical imaging
\citep{giavalisco04}, {\it Spitzer} IR to far-IR imaging (Dickinson
et~al., in prep; Chary et~al., in prep), {\it GALEX} far and near-UV
imaging \citep{schiminovich03}, ground-based optical and near-IR
imaging and spectroscopy \citep{capak04,cowie04,vanzella05}, and
radio/submillimeter observations (\citealt{richards00,pope05}).

Despite the easy access to broadband photometry and subsequent
insights into galaxy evolution made possible by multi-wavelength
surveys such as GOODS, important issues regarding survey completeness
and the physical conditions in galaxies and their surrounding
intergalactic medium can only be investigated spectroscopically.
Spectroscopy of galaxies in blind flux-limited surveys can be quite
inefficient and expensive, particularly if one only wants to study
galaxies at certain cosmological epochs.  However, we have shown that
the technique of photometric pre-selection can allow one to cull large
samples of galaxies in particular redshift ranges over a large range
in redshift $1.0\la z\la 4$ (e.g., \citealt{adelberger04,
steidel04,steidel03,steidel95,steidel93}), which can then be
efficiently followed up using multi-object optical spectrographs such
as the Low Resolution Imaging Spectrograph (LRIS) on the 10~m Keck
telescope.  Near-UV sensitive spectrographs such as the blue arm of
LRIS (LRIS-B) on Keck and the Focal Reducer/low dispersion
Spectrograph (FORS) on the VLT have significantly extended our
capabilities by allowing for spectroscopy of key features that fall
shortward of the OH emission forest for redshifts $1.4\la z\la 3$, a
particularly active epoch in the context of galaxy evolution and the
buildup of stellar and black hole mass.  To take advantage of
extensive multi-wavelength data, we included the GOODS-North (GOODS-N,
hereafter) field in our ongoing survey to select and spectroscopically
followup large samples of galaxies at redshifts $1.4\la z\la 3.0$
\citep{steidel04}.  In the interest of public dissemination of data,
we present in this paper the results of our spectroscopic survey of
$1.4\la z\la 3.0$ star-forming galaxies in the GOODS-N field including
associated photometry and spectroscopic redshifts.  Information on the
galaxies, including their photometric measurements and errors and
stellar population fits, are available at the following public
website: http://www.astro.caltech.edu/$\sim$drlaw/GOODS/.

The outline of this paper is as follows. In \S~\ref{sec:data} we
briefly describe the optical imaging, photometry, and spectroscopy.
To supplement these, we have also obtained the deepest wide-area
near-IR $J$ and $K$-band imaging in the GOODS-North field, and these
data are also presented in \S~\ref{sec:data}.  The spectroscopic
results and associated catalog are presented in \S~\ref{sec:specres}.
We describe the {\it Spitzer} IRAC and MIPS data (taken from the
GOODS-N public release; Dickinson et~al. in prep. and Chary et~al. in
prep.) for our spectroscopic sample of galaxies in
\S~\ref{sec:multiwave}.  Ground-based photometry and {\it Spitzer}
IRAC data, together with spectroscopic redshifts, enable the modeling
of the stellar populations of galaxies given certain simplifying
assumptions.  Our modeling procedure and results are discussed in
\S~\ref{sec:seds}.  In \S~\ref{sec:diversity} we describe a few
characteristics of the sample of star-forming galaxies and AGN to
demonstrate the wide range in intrinsic properties of UV-selected
galaxies at high redshift.  A flat $\Lambda$CDM cosmology is assumed
with $H_{0}=70$~km~s$^{-1}$~Mpc$^{-1}$ and $\Omega_{\Lambda}=0.7$.
All magnitudes are on the AB \citep{oke83} system.

\section{Data and Sample Selection}
\label{sec:data}

\subsection{Optical and Near-IR Imaging and Photometry}

The imaging, photometry, color selection, and spectroscopic
observations of galaxies in the fields of our $z\sim 2$ survey are
described in several other papers published by our group
\citep{steidel03,steidel04,adelberger04,reddy05a}.  Specific details
regarding the GOODS-N optical imaging are presented in
\citet{reddy05a} and summarized below for convenience.

The optical images used to photometrically pre-select candidate
galaxies at redshifts $1.4\la z\la 3.0$ in the GOODS-N field were
obtained in 2002 and 2003 April with the KPNO and Keck~I telescopes.
The KPNO MOSAIC $U$-band image obtained from the GOODS team (PI:
Giavalisco) was transformed to $U_{\rm n}$ magnitudes
\citep{steidel04}.  The Keck I $G$- and ${\cal R}$-band images were
taken with the LRIS instrument \citep{oke95,steidel04} and were
oriented to ensure maximum overlap with the GOODS {\it Spitzer} Legacy
and {\it Hubble} Treasury programs.  The images cover $11\arcmin
\times 15\arcmin$ with FWHM $\sim 0\farcs7-1\farcs1$ to a limiting
magnitude of $\sim 27.5$ measured within a $1\arcsec$ aperture
($3$~$\sigma$) in the $\ugr$ bands.  This depth ensures we are
complete to ${\cal R}=25.5$, neglecting photometric scatter, for
galaxies whose colors satisfy our $z\sim 2-3$ color criteria.  The
optical imaging reduction and photometry were done following the
procedures described in \citet{steidel03, steidel04}.  The images were
astrometrically calibrated using the SDSS database.  Source detection
is done at ${\cal R}$-band, and $\gmr$ and $U_{\rm n}-G$ colors are
computed by applying the $\rs$-band isophotal apertures to images in the
other filters.

\begin{figure*}[hbt]
\epsscale{0.8} 
\plotone{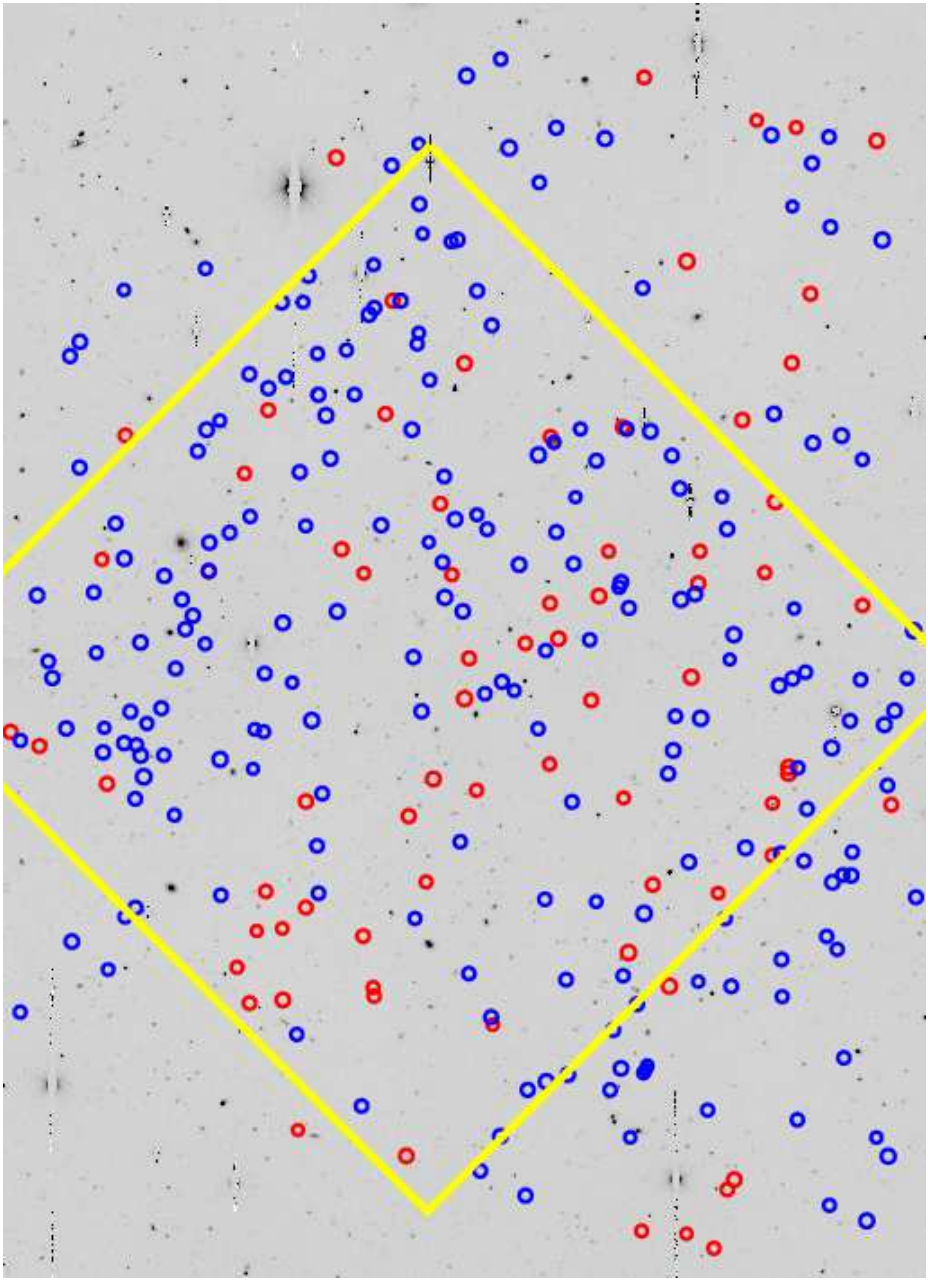}
\caption{Positions of BX/BM (blue circles) and LBG (red circles)
galaxies with spectroscopic redshifts $z>1.4$ overlaid on our
$10\arcmin \times 15\arcmin$ optical $\rs$-band image of the GOODS-N
field.  The yellow box ($8.5\arcmin \times 8.5\arcmin$) indicates the region 
with deep $J$ and $\ks$ {\it Palomar} imaging.
\label{fig:f1}}
\end{figure*}

We have obtained very deep wide-area near-IR imaging at $J$ and
$\ks$-band in the GOODS-N field from observations with the Wide Field
Infrared Camera (WIRC; \citealt{wilson03}) on the Palomar Hale $5$~m
telescope.  The images, taken under photometric conditions with $\sim
1\farcs0$ FWHM, reach a depth of $\ks\sim 24.4$ and $J\sim 25.0$ over
the central $8\farcm7~\times~8\farcm7$ of the GOODS-N field.  The
near-IR imaging reduction procedure is described in detail by
\citet{erb06b}.  Near-IR magnitudes were calibrated in Vega magnitudes
and converted to AB units assuming the following conversions: $\ks(AB)
= \ks(Vega)+1.82$ and $J(AB)=J(Vega)+0.90$.  Figure~\ref{fig:f1}
shows the area imaged in the near-IR with respect to our optical image
of the GOODS-N field.

Photometric errors for both optical and near-IR magnitudes were
determined from Monte Carlo simulations.  We added large numbers of
simulated galaxies with known magnitudes to our images and then
recovered them using the same photometric method used to detect actual
galaxies.  Comparing the input magnitudes with those recovered then
yields an estimate of the bias and uncertainty in our photometry.  The
Monte Carlo method is discussed in more detail by \citet{shapley05}.
The typical errors in the optical and near-IR magnitudes range from
$0.05$ to $0.3$~mag.

\subsection{Photometric Selection}

\begin{figure}[hbt]
\epsscale{1.0}
\plotone{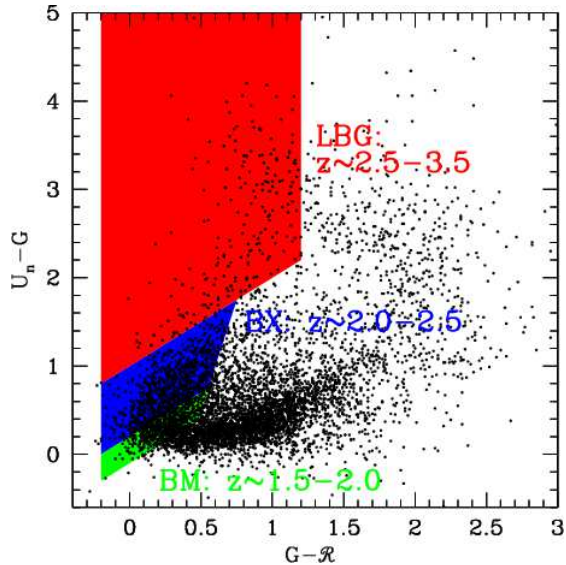}
\caption{$\ugr$ colors of objects with $\rs<25.5$ in the GOODS-N
field.  Also shown are the LBG, BX, and BM selection windows.
\label{fig:f2}}
\end{figure}

We selected galaxies in different redshift ranges between $1.4\la z\la
3.4$ using the ``BM'', ``BX'', ``C'', ``D'', and ``MD'' selection
criteria \citep{steidel03, adelberger04, steidel04}.  The ``C'',
``D'', and ``MD'' criteria are used to select Lyman Break Galaxies
(LBGs) at redshifts $2.7\la z\la 3.3$ \citep{steidel03}\footnote{Note
  that we did not select ``M'' galaxies in GOODS-N as was done in
  other fields of the $z\sim 3$ survey \citep{steidel03}.}.  The
``BM'' and ``BX'' criteria were designed to cull galaxies at redshifts
$1.4\la z\la 2.0$ and $2.0\la z\la 2.5$, respectively, with
approximately the same range of UV luminosity and intrinsic UV color
as the $z\sim 3$ LBGs \citep{adelberger04, steidel04}.  The various
selection criteria considered here are shown in Figure~\ref{fig:f2}.
We only considered candidates to ${\cal R}=25.5$ to ensure a sample of
galaxies which are bright enough such that optical spectroscopy is
feasible\footnote{A few objects were candidates based on photometry of
  our {\it Palomar} images of the GOODS-N field \citep{steidel03}, but
  failed to satisfy the photometric selection criteria based on the
  newer {\it Keck} images.  These objects are indicated in subsequent
  tables by their notation as presented in \citet{steidel03} or, in
  the case of BM/BX objects, by the letters ``BX'' or ``BM'' followed
  by no more than 3 numerical digits.  The photometric values for
  these objects are the ones based on the new photometry.}.  This
limit corresponds to an absolute magnitude $0.6$~mag fainter at $z\sim
2.2$ than at $z\sim 3$, based on the distance modulus between the two
redshifts.  We also excluded from the sample those candidates with
${\cal R}<19$ since almost all of these objects are stars.  Optical
selection yielded 1360 BM/BX and 192 C/D/MD candidates in the
$11\arcmin\times 15\arcmin$ area of the GOODS-N field.  Combined, the
BX/BM and C/D/MD candidates constitute $\sim 30\%$ of the ${\cal
  R}$-band counts to ${\cal R}=25.5$.  The number of candidates and
their surface densities are listed in Table~\ref{tab:counts}.
Approximately $50\%$ of these candidates lie in the region imaged at
$J$ and $\ks$.  The remainder of this paper focuses on those galaxies
which have been spectroscopically confirmed to lie at redshifts
$z>1.4$, as described in the next section.

\subsection{Optical Spectroscopy}

We took advantage of the multi-object capabilities of the Keck LRIS
instrument to obtain spectroscopy for the photometrically selected
candidates.  In its upgraded double-armed capacity, LRIS makes use
of a dichroic to send light to both a red and blue arm.
The commissioning of the blue arm of LRIS (LRIS-B)
allowed, for the first time, the ability to obtain very sensitive
near-UV spectroscopic observations at wavelengths as short as $\sim
3100$~\AA, essentially to the atmospheric transmission limit.  The
wavelength range from the atmospheric cutoff up to $\sim 5500$~\AA\,
is particularly useful for probing the rich set of interstellar and
stellar lines between Ly$\alpha$ and C~IV ($\lambda 1548,1550$) for
galaxies in the so-called ``spectroscopic desert'', between redshifts
$1.4\la z\la 2.5$.  As shown previously in \citet{adelberger04} and
\citet{steidel04}, combining photometric selection of BM and BX
candidates with the near-UV sensitivity of LRIS-B allows for the
wholesale spectroscopy of large numbers of galaxies in this redshift
range; this in turn enables us to focus our study on an epoch that was
particularly active in terms of star formation and accretion activity
(e.g., \citealt{dickinson03, chapman03, fan01, madau96, shaver96,
schmidt95}).

The instrumental setup used for spectroscopy in the GOODS-N field
varied during the course of the $z\sim 2$ survey; the various setups
are described in more detail by \citet{steidel04}.  We selected
dichroic filters designed to split the incoming beam at $5600$~\AA\,
or $6800$~\AA.  To provide maximum throughput between $3100$ and
$4000$~\AA, we used a $400$~groove~mm$^{-1}$ grism blazed at
$3400$~\AA\, on LRIS-B, resulting in a dispersion of $1.09$~\AA\, per
pixel.  For simultaneous observations on the red-side of LRIS
(LRIS-R), we used a $400$~groove~mm$^{-1}$ grating blazed at
$8500$~\AA, providing wavelength coverage up to $9500$~\AA.  We
typically obtained simultaneous blue and red side spectroscopic
observations between $3100$ and $9500$~\AA, with slight variations due
to the relative placement of slits in the telescope focal plane.

The slit masks used for spectroscopy cover $8\arcmin\times 5\arcmin$
on the sky.  For a minimum slit length of $9\arcsec$ (adopted in order
to ensure good background subtraction), we are able to include $30-35$
slits, in addition to $4-5$ star boxes used to accurately align each
mask.  We set the width of each slit to $1\farcs2$ and this, combined
with a typical seeing of $0\farcs8$, yields a typical resolution of
$5$~\AA\, for point sources.  To obtain the optimum mix of objects on
any given slit mask, we assigned each candidate a weight primarily
based on its optical magnitude.  We gave larger weights to objects
with ${\cal R}=23.5-24.5$ and lower weights for fainter objects where
absorption line spectroscopy is more difficult and brighter objects
where the foreground ($z\la 1.0$) interloper fraction is larger.
Nonetheless, we filled ``blank'' areas of the masks with filler
objects which included these fainter and brighter objects.  In
particular, we included some bright (${\cal R}<23.5$) objects on masks
since at least some of these are intrinsically bright $z\sim 2-3$
galaxies and are most suitable for detailed followup spectroscopic
studies.  To support other projects being conducted by our group, we
also deliberately targeted objects within the $\ugr$ sample that had
interesting multi-wavelength properties, such as those identified with
$850$~$\mu$m or $24$~$\mu$m emission, as well as those with unusually
red near-IR colors.  We also designed masks to overlap as much as
possible with the near-IR imaging.  Because of this, $\sim 73\%$ of
spectroscopically confirmed galaxies with $z>1.4$ lie in the
$\ks$-band region, even though $\sim 50\%$ of $\ugr$ candidates lie in
the same region (see Figure~\ref{fig:f1}).

We typically obtained 3 exposures of $1800$~sec per mask, for a total
exposure of $5400$~sec.  The range in optical magnitudes implies a
large range in the S/N of the spectra.  At the minimum, however, we
found $5400$~sec to be sufficient to obtain redshifts, and a few
objects were observed on more than one mask.  The spectroscopic
success rate per mask is primarily a function of the weather
conditions (e.g., cirrus, seeing) at the time of observation, with a
$90\%$ success rate of obtaining redshifts in the best conditions;
these redshifts, for the most part, fell within the targeted redshift
ranges.  This suggests that the redshift distribution for the
spectroscopic and photometric samples are similar, and there are not
large numbers of galaxies whose true redshifts are far from those
expected based on their observed $\ugr$ colors.  Details of the
spectral reduction techniques are described in \citet{steidel03}.  We
identified redshifts based on the presence of a number of
low-ionization interstellar absorption lines (e.g., Si~II $\lambda
1260$, O~I+S~II $\lambda 1303$, C~II $\lambda 1334$, Si~II $\lambda
1526$, Fe~II $\lambda 1608$, Al~II $\lambda 1670$, and Al~III $\lambda
1854,1862$), stellar wind features (e.g., N~V $\lambda 1238,1242$,
S~IV $\lambda 1393,1402$, and C~IV $\lambda 1548, 1550$), the C~III
$\lambda 1909$ nebular emission line, or Ly$\alpha$ emission or
absorption.  A few examples of spectra are shown in \citet{steidel03,
steidel04}.  Spectroscopically confirmed galaxies with $1.4<z<3.0$ are
shown with respect to the $\rs$-band image of the GOODS-N field in
Figure~\ref{fig:f1}.

Comparison with nebular redshifts derived from H$\alpha$ spectroscopy
indicates that Ly$\alpha$ emission is almost always redshifted, and
interstellar absorption lines are almost always blueshifted, with
respect to the systemic (nebular) redshift of the galaxy.  These
systematic offsets have been interpreted as the result of outflows
(e.g., \citealt{adel03, adelberger05b, pettini01, shapley03}).
\citet{adelberger05b} present linear least-squares fits to the
systemic redshifts of galaxies given their Ly$\alpha$ and interstellar
absorption redshifts based on a sample of $138$ objects with near-IR
spectroscopy \citep{erb06b, pettini01}.

\section{Spectroscopic Results and Catalog}
\label{sec:specres}

Our spectroscopic sample in the GOODS-N field presently includes 212
BM/BX and 74 C/D/MD galaxies with secure spectroscopic redshifts
$z>1.4$ (Table~\ref{tab:galaxies}).  The total sample includes 347
objects with secure spectroscopic redshifts, including $40$
interlopers with $z<1$.  We also include 41 objects with uncertain
redshifts in Table~\ref{tab:galaxies}, denoted by a colon (``:'') in
the redshift field (for consistency with \citealt{steidel03}), for a
total of $388$ objects.  Table~\ref{tab:counts} lists the statistics
for the individual samples, including the numbers of candidates
observed, the interloper fractions, and mean redshifts.  The primary
source of contamination in the LBG sample is from K dwarfs in the
Galactic halo.  Star-forming galaxies at redshift $\langle z\rangle =
0.17\pm 0.09$ contaminate the BX sample since their Balmer breaks
mimic the Ly$\alpha$ forest decrement.  These interlopers can be
easily excluded using $\ks$-band photometry (e.g., the $\bzk$
criteria; \citealt{daddi04}), but we have not imposed any additional
criteria other than the observed optical magnitudes and colors.  The
main ``contaminants'' of the BM sample occur from galaxies with
redshifts $1.0<z<1.4$; these galaxies have $\ugr$ colors very similar
to those of BM objects, and the narrow BM color selection window
implies that photometric scatter and Ly$\alpha$ perturbations on the
$\ugr$ colors can have a significant impact on the observed redshift
distribution of BM galaxies (Reddy et~al., in prep.).  Throughout this
paper we consider objects with $z<1.4$ to be contaminants.  The
AGN/QSO (as identified from either their X-ray, UV, or {\it Spitzer}
IRAC or MIPS emission) with $z>1.4$ make up $\sim 4\%$ of the sample
(see \S~\ref{sec:agn} for further discussion).  

For consistency, we compared redshifts for objects in common with the
Team Keck Treasury Redshift Survey (TKRS; \citealt{wirth04, cowie04}).
We note that the TKRS survey is based primarily upon observations with
the DEIMOS intrument on Keck~II, and so the TKRS redshift selection
function rapidly declines above $z\sim 1.2$ as the emission and
absorption lines used for redshift identifiction (including [N~II],
[S~II], [O~III], [O~II] emission features and Calcium H and K
absorption features) are shifted out of the DEIMOS spectral range.
The overlap between the $\ugr$ and TKRS samples is small given that
the two surveys target different redshifts (TKRS is better at
identifying galaxies at $z\la 1.2$ and our $\ugr$ selection is better
at identifying galaxies at $z\ga 1.4$).

There are 64 objects with redshifts in the $\ugr$ catalog which are
also in the TKRS database.  Of these, 52 were previously published in
other surveys of the GOODS field \citep{cowie04, cohen96, cohen00,
cohen01, barger00, barger03, wirth04, phillips97, lowenthal97,
dawson01, steidel03, steidel96, dickinson98} and/or have agreement in
redshift between the $\ugr$ and TKRS samples.  Upon further inspection
of the 12 objects with discrepant redshifts, we adopted the TKRS
redshift for 6 of them (BX1202, BX1371, BMZ1010, BMZ1100, BMZ1121, and
BMZ1208); the redshifts for these 6 galaxies are all below $z<1.4$
where the DEIMOS-determination was found to be secure and where our BM
selection function drops off.  Five of the objects had the correct
redshifts in our catalog (BX1299, BX1319, BX1805, BMZ1119, and
BMZ1375).  For the remaining object, BX1214, we were able to rule out
the \citet{cohen00} redshift of $z=2.500$, but were unable to
confidently assign a redshift based on our LRIS spectrum.

\begin{figure*}[hbt]
\plottwo{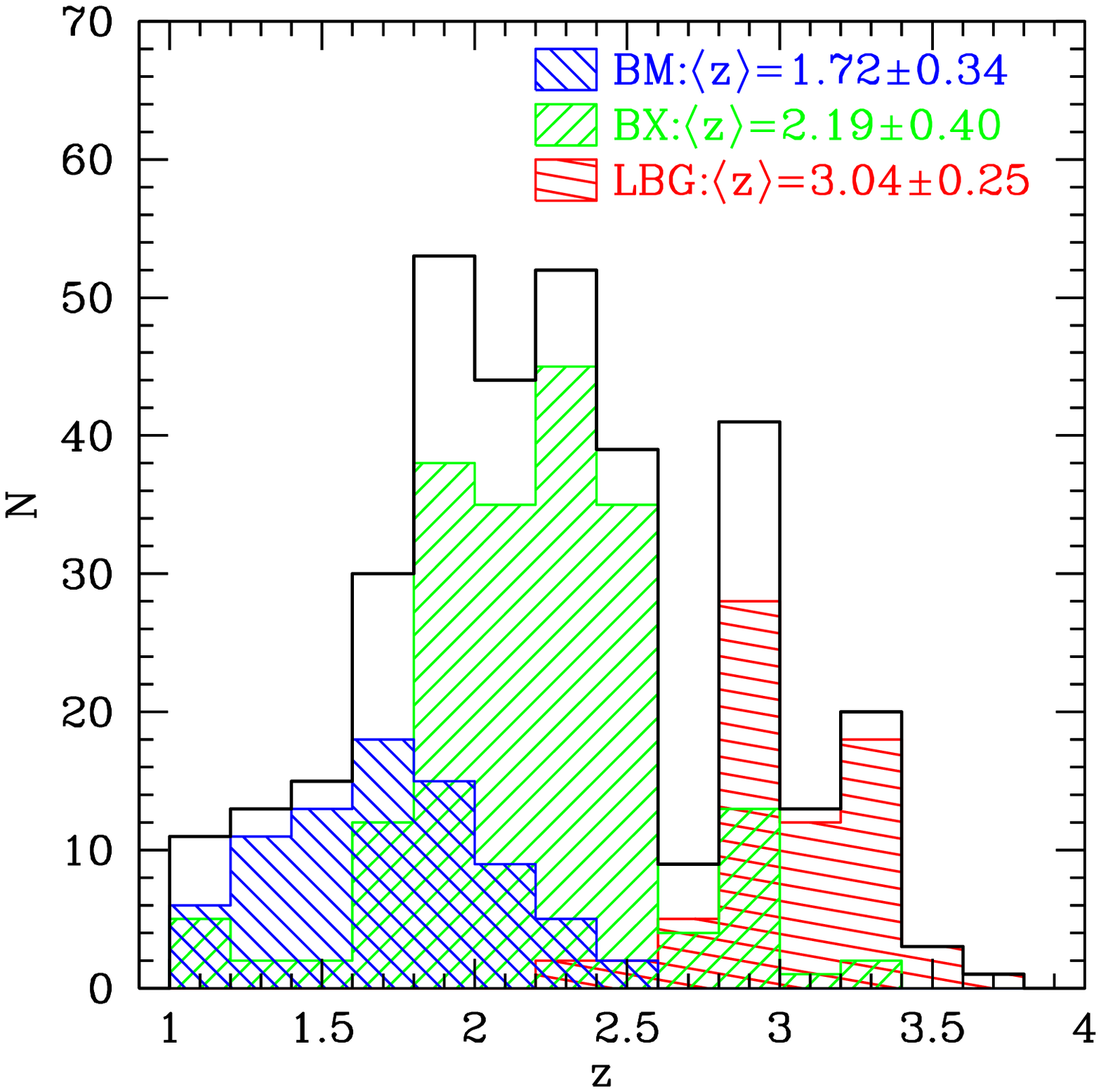}{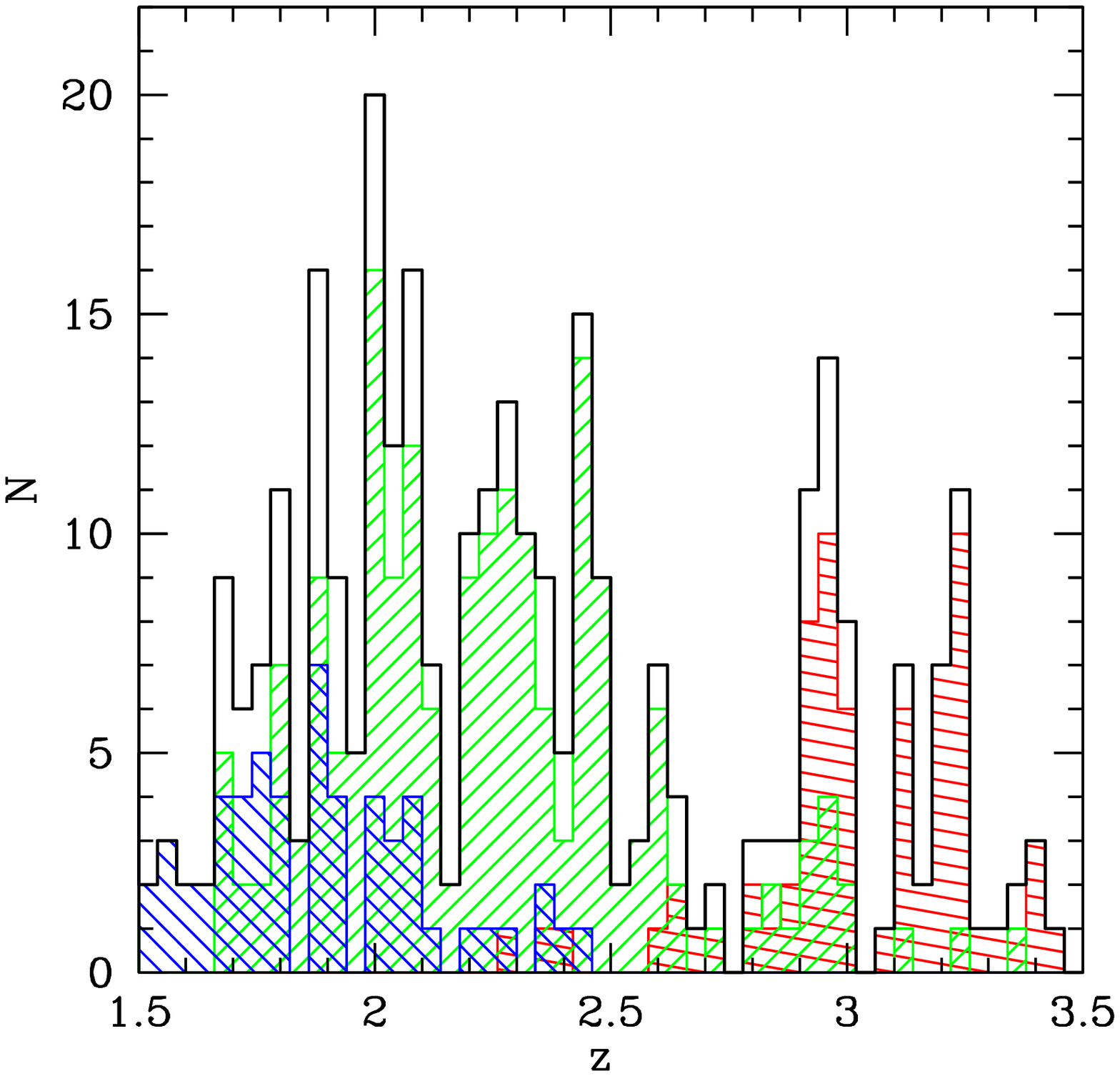}
\caption{({\it Left}) Redshift histogram of spectroscopically
confirmed BM/BX and LBG galaxies in the GOODS-N field with $z>1$.  The
solid line indicates the total distribution of BM, BX, and LBG
galaxies. ({\it Right}) Redshift histogram with higher resolution
bins, $\delta z = 0.04$, emphasizing large-scale structure in GOODS-N.
\label{fig:f3}}
\end{figure*}

The redshift distributions of BM/BX and LBG galaxies with $z_{\rm
spec}>1$ are shown in Figure~\ref{fig:f3}, where the right panel
emphasizes large-scale structure in the GOODS-N field.  The redshift
over-density at $z=2.95$ is prominent, and was also noted in the LBG
survey \citep{steidel03}.  We also note a possible over-density at
$z=2.00$, which corresponds to an over-density of 5 submillimeter
galaxies as noted by \citet{blain04}.  However, we caution that
Figure~\ref{fig:f3} only presents raw numbers, and we have not
accounted for the selection function and relative fractions of
candidates observed.  Therefore, the significance of any redshift
``over-densities'' appearing in Figure~\ref{fig:f3} is not quantified.

\section{{\it Spitzer} IRAC and MIPS Data}
\label{sec:multiwave}

To aid in understanding the stellar populations and extinction of
$z\sim 2$ galaxies, we compiled {\it Spitzer} IRAC and MIPS photometry
for our sample of $\ugr$-selected galaxies using the public {\it
Spitzer} data in the GOODS-N field (Dickinson et~al., in prep; Chary
et~al., in prep).  The IRAC photometry was performed by fitting an
empirical point spread function (PSF) determined from the IRAC images
to the spatial positions of sources from the higher resolution
$\ks$-band data.  Specifically, to accurately compute the empirical
PSF for each IRAC channel we used as many isolated point sources
located throughout the IRAC images as possible.  Changing the number
of objects used to compute the PSF results in variations of the
best-fit fluxes of $<5\%$, and is generally small compared to the
background noise (see below).  Using the $\ks$-band data to constrain
source positions mitigates the effects of confusion by allowing for
the deblending of partially resolved sources in the IRAC images, and
is similar to the method employed by the GOODS team for extracting
photometry (Dickinson et~al., in prep; Chary et~al., in prep; see also
\citealt{labbe05} and \citealt{shapley05}).  We extracted the MIPS
$24$~$\mu$m fluxes of $\ugr$ galaxies using a similar procedure; the
spatial positions of sources from the IRAC data were used to deblend
and extract the $24$~$\mu$m fluxes \citep{reddy06a}.  Errors for both
IRAC and MIPS photometry were computed from the dispersion of
extracted fluxes for 100 PSFs fit to random blank regions around each
galaxy.  Since the IRAC and MIPS data are background-limited, the
errors will be dominated by the background noise for all but the
brightest galaxies at these wavelengths.  The IRAC channel $1-4$
magnitudes and MIPS $24$~$\mu$m fluxes are listed in
Table~\ref{tab:galaxies}.\footnote{Magnitude and flux errors for the
{\it Spitzer} data are provided in the accompanying machine-readable
table.}  We do not give fluxes for those galaxies
which were either undetected or were badly blended with a nearby
bright source.  Of the 212 BX/BM galaxies with secure spectroscopic
redshifts $z>1.4$, only 2 ($<1\%$) are undetected at $3.6$~$\mu$m to
the GOODS IRAC depth.  Of the 74 LBGs, 11 ($\approx 15\%$) are
undetected at $3.6$~$\mu$m.  The MIPS detection fraction is $\approx
65\%$ for BX/BM galaxies, decreasing to a $\approx 53\%$ for the LBGs,
to a limiting $3$~$\sigma$ flux of $f_{\rm 24\mu m} \approx
8$~$\mu$Jy.

\section{Stellar Population Modeling}
\label{sec:seds}

The combination of multi-wavelength photometry and spectroscopic
redshifts allows us to better constrain the stellar populations of
UV-selected galaxies than if we only had photometric redshifts.  To
demonstrate the wide range in stellar populations of UV-selected
galaxies at redshifts $z\sim 2-3$, we fit the $\ugr J\ks$ $+$ IRAC
magnitudes with \citet{bruzual03} models assuming a \citet{salpeter55}
IMF and solar metallicity.  The assumption of solar metallicity is a
reasonable approximation for most galaxies in the $\ugr$-selected
sample \citep{erb06a}.  The models were corrected for the effects of
IGM opacity before comparing to the observed magnitudes.  In fitting
the stellar populations, we assumed an exponentially-declining star
formation history with decay time scales $\tau$ $=$ 10, 20, 50, 100,
200, 500, 1000, 2000, and 5000 Myr, and $\tau=\infty$ (constant star
formation, CSF, model).  We also assumed a varying amount of
reddening, or $\ebmv$, from $0.0$ to $0.7$.  The best-fit model was
taken to be the combination of $\tau$, age, and $\ebmv$ that gave the
lowest $\chi^2$ value with respect to the observed magnitudes.  The
star formation rate (SFR) and stellar mass are determined from the
normalization of the model to the observed magnitudes.  Even with
spectroscopic redshifts, there is considerable uncertainty in the
best-fit parameters, with the exception of the total stellar mass
$M^{\ast}$ which is generally robust to changes in the assumed star
formation history (e.g., \citealt{papovich01, shapley01, sawicki98,
shapley05, erb06b}).  The best-fit stellar population parameters for
both a CSF and $\tau$ model for each galaxy are collected in
Table~\ref{tab:sedparms}.  Monte Carlo simulations indicate that the
typical fractional uncertainties associated with the best-fit
parameters (when including IRAC data in the fits) are $\langle
\sigma_x/\langle x\rangle\rangle = 0.6$, $0.4$, $0.5$, and $0.2$ in
$\ebmv$, age, SFR, and stellar mass, respectively \citep{erb06b}.  For
completeness we have included the best-fit SFRs from the fitting in
Table~\ref{tab:sedparms}, but we note that we have several other {\it
independent} multi-wavelength measures of the SFRs for these galaxies
(e.g., from dust-corrected UV, H$\alpha$, and $24$~$\mu$m data) that
are unaffected by the degeneracies associated with stellar population
modeling.

Aside from the systematic errors resulting from the degeneracy between
star formation history and the best-fit parameters, there are
additional caveats to the SED results.  Around 30 objects had optical
through IRAC photometry which is inconsistent with the stellar
population models considered here; these objects exhibit large J/$\ks$
{\it and} IRAC magnitude residuals with respect to the best-fit
stellar population (and have $\chi^2>10$), and often give
unrealistically young ages ($<10$~Myr) and large SFRs
($>2000$~$\sfr$).  We do not present the SED results for these
galaxies.  In addition to these 30, there are 4 galaxies which fit the
optical and IRAC data well, but have large $\ks$ residuals with
respect to the best-fit stellar population (i.e., a $\ks$ magnitude
more than $3$~$\sigma$ away from the best-fit).  Three of these 4
galaxies have redshifts $2.0\le z\le2.5$ where the $\ks$ magnitude may
be contaminated by emission from H$\alpha$+[N\,{\sc ii}].  The 4
galaxies with large $\ks$ residuals are indicated by the notation
``$\ks$'' in Table~\ref{tab:sedparms}.  Also, we noted a few objects
with $8$~$\mu$m excesses when compared with the best-fit stellar
population, many of which have large $24$~$\mu$m fluxes ($f_{\rm 24\mu
m}>100$~$\mu$Jy) indicating they may be obscured AGN (see
Table~\ref{tab:agn}).  We do not present SED fitting results for any
of the sources which may have AGN based on their $8$~$\mu$m and
$24$~$\mu$m excesses and/or X-ray/optical emission.  Finally, we did
not perform SED fitting for galaxies without photometry longward of
${\cal R}$-band or which had redshifts $z<1$.  The best-fit SED
parameters for the remaining $254$ galaxies are listed in
Table~\ref{tab:sedparms}.  Note that the SFRs and stellar masses
($M^{\ast}$) in Table~\ref{tab:sedparms} assume a \citet{salpeter55}
IMF from $0.1$ to $100$~M$_{\odot}$.  Assuming the \citet{chabrier03}
IMF with a shallower faint-end slope results in SFRs and stellar
masses a factor of $1.8$ lower than listed in
Table~\ref{tab:sedparms}.  We also note that a number of galaxies have
inferred ages $<50$~Myr, which are unlikely given the dynamical
time-scale of $\sim 50$~Myr for star formation in galaxy-sized
objects.  The SED parameters for these galaxies with extremely young
{\it inferred} ages should be taken with caution.

\section{The Diverse Properties of Optically-Selected Galaxies at
High Redshift}
\label{sec:diversity}

\subsection{Star-Forming Galaxies}

\begin{figure}[hbt]
\plotone{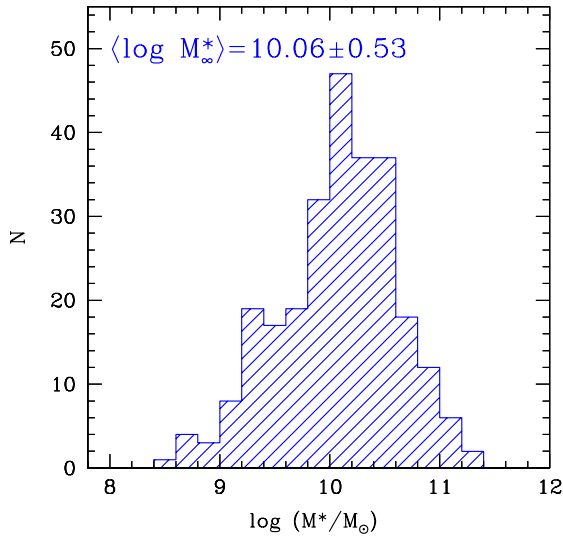}
\caption{Stellar mass distribution of $\ugr$-selected galaxies with
redshifts $z>1$, assuming a constant star formation (CSF) model.
Assuming a best-fit exponentially declining star formation history
($\tau$ model) results in a stellar mass distribution that is
virtually identical to the one shown here, with a mean and dispersion
in log space of $\langle \log M^{\ast}_{\tau}\rangle = 10.08\pm0.51$.
\label{fig:f4}}
\end{figure}

Of the best-fit SED parameters, the stellar mass is the least
uncertain and is generally robust to changes in the assumed star
formation history, as can be seen by comparing columns 6 and 11 in
Table~\ref{tab:sedparms}.  Figure~\ref{fig:f4} shows the
distribution of stellar masses for $\ugr$-selected galaxies with
redshifts $z>1$, assuming a constant star formation history.
Table~\ref{tab:masses} shows the median and mean stellar masses and
dispersion (assuming a CSF history) for galaxies in the various
samples.  While the mean stellar mass of the sample is $\langle
M^{\ast}\rangle \approx 1.1\times 10^{10}$~M$_{\odot}$, there is large
dispersion about this mean of a factor of $3.4$.  This mean stellar
mass is a factor of $\sim 2$ lower than found in \citet{shapley05} and
\citet{erb06b}, partly because we included galaxies undetected to
$\ks=24.1$ in the sample considered here (as long as they had IRAC
data to constrain the stellar mass), and these faint $\ks$ galaxies on
average have lower stellar masses than $\ks$-detected galaxies.
Further, we have included BM galaxies which have a mean stellar mass
that is a factor of $\approx 2$ lower than than the mean stellar mass
for BX galaxies and LBGs (Table~\ref{tab:masses}).  This difference in
mean stellar mass likely reflects the fact that BM galaxies have a
lower mean redshift ($z=1.72\pm0.32$; Figure~\ref{fig:f3}) than BX
galaxies and LBGs, and therefore we are able to probe down to fainter
absolute magnitudes and are sensitive to lower mass galaxies.

Regardless, the sample includes galaxies with a wide range in ages,
from young galaxies with ages comparable to the dynamical timescale
for star formation of $\sim 50$~Myr to those which are older than
$2$~Gyr.  In fact, the $\ugr$ sample includes galaxies which are as
old ($>2$~Gyr) and as massive ($M^{\ast}>10^{11}$~M$_{\odot}$) as
galaxies found at $z\sim 2$ in near-IR selected samples (e.g., the
distant red galaxies, or DRGs, of
\citealt{franx03})\footnote{Optically selected galaxies which satisfy
the DRG criteria ($\jmk>2.3$ in Vega magnitudes, or $\jmk>1.38$ in AB)
are indicated by ``DRG'' in the last column of
Table~\ref{tab:galaxies}.  There are $19$ such DRGs with secure
spectroscopic redshifts in our sample.}.  In particular,
\citet{shapley05} and \citet{reddy05a} have shown that while the
typical stellar mass of near-IR selected DRGs with $\ks\la 21.8$ is
larger by an order of magnitude than the {\it typical} stellar mass of
$\ugr$-selected galaxies to ${\cal R}=25.5$, the actual range in
stellar mass probed by DRG selection does not appear to significantly
exceed the range in stellar mass of $\ugr$-selected galaxies (although
we note that DRG selection appears to be much more efficient is
selecting galaxies with $M^{\ast}\ga 10^{11}$~M$_{\odot}$ at $z\sim
2-3$; e.g., \citealt{vandokkum06}).  Further, SED analysis of the
optically-selected DRGs (as indicated in Table~\ref{tab:galaxies} by
the notation ``DRG'' in the last column) with fainter near-IR
magnitudes ($\ks\ga 22.8$) have stellar masses that are comparable to
the stellar masses of typical $\ugr$-selected galaxies ($10^{9}\la
M^{\ast}\la 10^{11}$~M$_{\odot}$).  While optical selection allows us
to very efficiently followup galaxies spanning over two orders of
magnitude in age and stellar mass at redshifts $z\ga 1.4$, other
techniques are required to assess the total stellar mass budget at
these redshifts (e.g., \citealt{rudnick06, vandokkum06, reddy05a}).

While the $\ebmv$ and SFRs determined from SED modeling are more
uncertain than the inferred stellar masses, we have several
independent methods of assessing the extinction and SFRs in $z\sim 2$
galaxies, made possible by the extensive multi-wavelength data in the
GOODS-N field.  The exquisite, photon-limited {\it Chandra} X-ray data
in the GOODS-N field, currently the deepest X-ray data ever taken
\citep{alexander03}, allow for stacking analyses to estimate the
average emission properties of galaxies \citep{brandt01, nandra02,
  reddy04, reddy05a}.  Based on the stacking analyses of
\citet{nandra02} and \citet{reddy04}, the mean SFRs of $z\sim 2-3$
$\ugr$-selected galaxies is $\sim 50$~$\sfr$, with mean attenuation
factors, defined as the ratio between the bolometric SFR and UV-based
SFR (uncorrected for extinction), of $4.5-5.0$.

The X-ray data allow us to determine the average extinction and SFRs
of galaxies over the entire range of redshifts probed by the BX/BM and
LBG criteria.  However, important progress has been made in
determining the individual properties of galaxies in a narrower
redshift range, $1.5\la z\la 2.6$, where the {\it Spitzer} MIPS
$24$~$\mu$m band is sensitive to the $7.7$~$\mu$m polycyclic aromatic
hydrocarbon (PAH) dust emission ubiquitous in local and high redshift
star forming galaxies \citep{reddy06a, papovich06}.  \citet{reddy06a}
demonstrate that the $24$~$\mu$m emission of $z\sim 2$ galaxies can be
used as a tracer of the SFR or total infrared luminosity ($L_{\rm
IR}$), particularly for galaxies with spectroscopic redshifts where we
are able to accurately constrain the {\it K}-corrections from
$24$~$\mu$m flux to rest-frame $5-8.5$~$\mu$m luminosity.  The MIPS
data indicate that $\ugr$-selected galaxies at redshifts $1.5\la z\la
2.6$ span more than 3 orders of magnitude in $L_{\rm IR}$, from those
which are undetected to the $3$~$\sigma$ sensitivity limit of
$8$~$\mu$Jy for MIPS data in the GOODS field, to those which have
$L_{\rm IR}$ comparable to the most luminous star-forming galaxies at
these redshifts, the submillimeter galaxies \citep{smail97, hughes98,
barger98, chapman05}.  The mean infrared luminosity for
$\ugr$-selected galaxies is $\langle L_{\rm IR}\rangle \simeq 2\times
10^{11}$~L$_{\odot}$, assuming that the rest-frame infrared emission
($L_{\rm 5-8.5\mu m}$) as probed by MIPS observations scales with
infrared luminosity as $L_{\rm IR}\approx 17.2 L_{\rm 5-8.5\mu m}$ as
determined from local samples (see \citealt{reddy06a}), and this value
of $\langle L_{\rm IR}\rangle$ inferred from MIPS is in excellent
agreement with X-ray and dust-corrected UV-based estimates.
\citet{reddy06a} further demonstrate that the agreement in SFR extends
to galaxies on an individual (object-by-object) basis for a small
sample of $\ugr$-selected galaxies with both $24$~$\mu$m and H$\alpha$
detections: the scatter between the bolometric luminosity inferred
from $24$~$\mu$m versus H$\alpha$ observations is $\sim 0.2$~dex.
Finally, we have compared the $24$~$\mu$m-estimated SFRs with those
obtained from the SED-fitting analysis: as expected they are
positively correlated at the $5$~$\sigma$ level with a scatter of
$0.3$~dex assuming a CSF model.\footnote{Adopting a declining model,
as opposed to assuming a CSF model for all galaxies, will generally
increase the scatter in SFRs since the youngest galaxies will have
inferred SFRs (from declining star formation history models) that are
systematically lower than those inferred from H$\alpha$ observations
(see discussion in \citealt{erb06c}).}  The advantage of
multi-wavelength data is that we can assess SFRs independent of the
degeneracies associated with SED-fitting.  In summary, the
$\ugr$-selected sample includes galaxies over 4 orders of magnitude in
dust obscuration ($L_{\rm bol}/L_{\rm UV}$), from those galaxies with
little dust and whose UV luminosity is comparable to $L_{\rm IR}$, to
those which are heavily dust-obscured and have attenuation factors
$\ga 1000$.

Aside from the large dynamic range in SFRs and extinction of
$\ugr$-selected galaxies, the sample also hosts galaxies with a wide
range in morphology and kinematics (\citealt{erb03, erb06b}; Law
et~al., in prep.), from disk-like galaxies with signatures of
rotation, as inferred from H$\alpha$ spectral data (e.g.,
\citealt{forster06}), to those galaxies which appear irregular and/or
are merging.  UV-selected samples efficiently target the redshift
range where the morphological transformation of galaxies from
irregular at high redshift to the Hubble sequence at low redshifts
($z\la 1.4$) takes place.  The deep Hubble ACS data in the GOODS field
\citep{giavalisco04} combined with our extensive rest-frame UV
spectroscopic database make it possible to study in detail the
correlation between morphological structure and the SFRs, extinction,
masses, and spectral properties of high redshift galaxies (Law et~al.,
in prep.).

\subsection{AGN}
\label{sec:agn}

The combination of X-ray, observed optical, $8$~$\mu$m, and
$24$~$\mu$m data, along with spectroscopic redshifts, allows for a
powerful probe of AGN activity among $\ugr$-selected galaxies.  We
classified objects as AGN based on one or more of the following
criteria: (a) the presence of high ionization UV lines (identical to
the method used in \citealt{steidel02} and \citealt{shapley05}); (b)
direct detection in the {\it Chandra} 2~Ms data \citep{alexander03}
and an X-ray-to-optical flux ratio indicative of AGN (e.g., see
\citealt{hornschemeier01} and \citealt{reddy05a}); or (c) an $8$ and
$24$~$\mu$m flux excess above what one would expect from a simple
star-forming population.  Table~\ref{tab:agn} lists the 11 AGN with
confirmed redshifts $z>1.4$ which have emission indicative of AGN.

MD31 is the most unusual source: it has an X-ray counterpart within
$1\farcs 5$ of the optical position, but shows no evidence of AGN from
the rest-UV (observed optical) spectrum nor from {\it Spitzer}
observations.  The SED analysis indicates that MD31 is best-fit with
an $\sim 2$~Gyr old population with a modest $\ebmv\sim 0.17$ and
SFR$\sim 60$~$\sfr$ assuming the CSF model, and thus is not expected
to be bright in X-rays as a result of star formation alone (i.e., the
2~Ms X-ray sensitivity implies a detection threshold of $\sim
480$~$\sfr$ at the redshift of MD31, $z=2.981$).  Examination of the
deep ACS imaging in the GOODS field reveals no other optical
counterpart within $1\farcs 5$ of MD31.  If the X-ray counterpart is
indeed associated with accretion activity in MD31, then the X-ray
detection fraction of AGN with $z>1.4$ in our sample is $7/11$, or
$64\%$.  On the other hand, the fraction of AGN showing $8$~$\mu$m
and/or $24$~$\mu$m excesses is $9/11$, or $82\%$.  While the object
statistics are insufficient to judge the efficiency of AGN detection
in the X-ray versus IR, we note that the IRAC and MIPS integration
time for any given object in the GOODS-N field is $\sim 10$~hours,
whereas the X-ray integrations required to detect faint AGN at
redshifts $z\ga 1.4$ is on the order of a megasecond or larger.  The
possible difference in AGN detection fraction and especially
integration time between the IR and X-ray observations suggest that
deep IR imaging may be a more efficient method of finding AGN at high
redshifts.  In this case, the $8$~$\mu$m and $24$~$\mu$m data indicate
at least an additional 3 AGN which are unidentified in X-rays.  For
comparison, while our optical spectra have integration times of
$5400$~s, a factor of 370 times shorter than the X-ray integration
time ($2$~Ms), we can still detect $\approx 73\%$ of AGN based on
their rest-frame UV emission lines.

\begin{figure*}[hbt]
\plotone{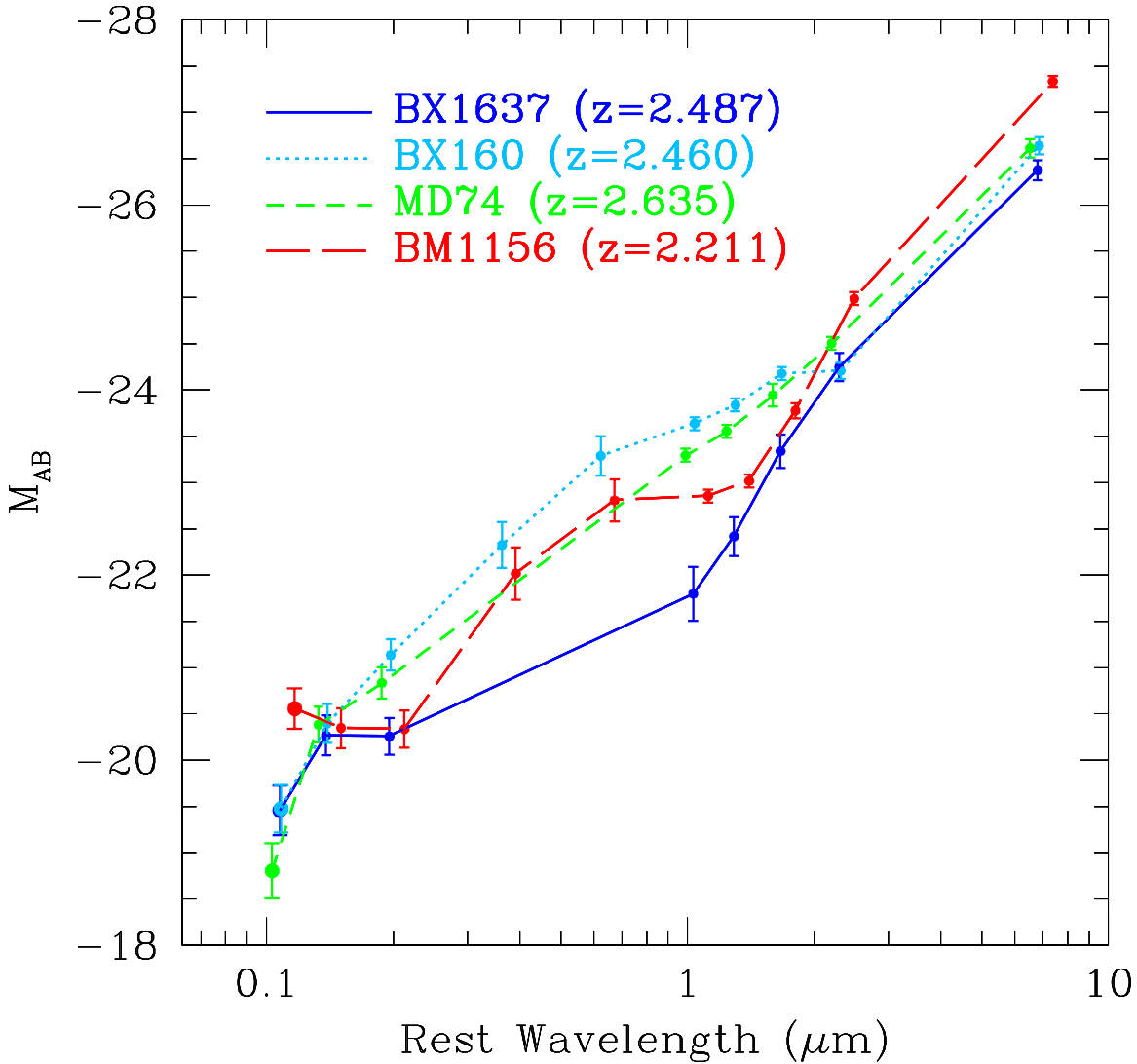}
\caption{Spectral energy distributions of the three X-ray undetected
AGN (BX1637, BX160, and MD74) and one faint X-ray detected AGN
(BM1156) at $z=2.211-2.635$ in the spectroscopic $\ugr$ sample, from
observed optical through $24$~$\mu$m.  All four exhibit a power law
slope at long wavelengths ($\lambda\ga 2$~$\mu$m rest-frame)
indicative of a warm dust population.
\label{fig:f5}}
\end{figure*}

The properties of the 3 X-ray undetected AGN with $z>2$ are worth
further consideration: these AGN are BX1637, BX160, and MD74.
Figure~\ref{fig:f5} shows the observed optical through
$24$~$\mu$m SEDs of these AGN, as well as BM1156 which is very weakly
detected in X-rays, and demonstrates the power-law behavior at
observed wavelengths longer than $\lambda\sim 2$~$\mu$m, indicative of
warm dust population.  The three X-ray undetected AGN are also not
detected in the deep radio imaging of the HDF \citep{richards00},
placing a $3$~$\sigma$ upper limit on their observed $1.4$~GHz flux of
$f_{\rm 1.4~GHz}<24$~$\mu$Jy.  However, all three have disturbed
and/or extended rest-frame UV morphologies from deep ACS imaging
\citep{giavalisco04}, suggesting the obscured AGN in these systems may
be triggered by merger activity.  The non-detection of these three AGN
(even when stacking them) in both the soft ($0.5-2.0$~keV) and full
($2.0-8.0$~keV) X-ray bands of {\it Chandra} makes it difficult to
constrain their column densities.  Nonetheless, these three AGN have a
mean spectroscopic redshift of $\langle z_{AGN}\rangle\approx 2.5$,
and if we assume the AGN have intrinsic photon index of $\Gamma=2.0$
(e.g., \citealt{alexander05}), then we cannot rule out the possibility
that these could be Compton-thick AGN with column densities $N_{\rm
H}>10^{24}$~cm$^{-2}$.  Comparison with the $8$~$\mu$m and $24$~$\mu$m
fluxes of Mrk~463 and Mrk~1014, two infrared luminous AGN
\citep{armus04}, suggests that BX1637 and MD74 could have total
infrared luminosities in the range $5\times 10^{12} \la L_{\rm IR}\la
2\times 10^{13}$~L$_{\odot}$.  Constraints on the $850\mu$m fluxes of
these two AGN could narrow the range of possible $L_{\rm IR}$, but
unfortunately submillimeter observations do not cover the region
containing these two AGN.  BX160 is covered in published $850$~$\mu$m
imaging and is undetected to $S_{\rm 850}\sim 1.5$~mJy \citep{wang04},
suggesting an upper limit on the infrared luminosity of $\sim
10^{12.5}$~L$_{\odot}$.  We caution, however, that the lack of data
across the Raleigh-Jeans tail of the dust SEDs makes it difficult to
accurately constrain the dust temperatures and hence total bolometric
luminosities of these sources.

The presence of obscured AGN at high redshifts has been postulated
based on the expected fraction of high column density AGN ($N_{\rm
H}>10^{23}$~cm$^{-2}$) at $z\ga 1.4$ in simulations that model the
contribution to the X-ray background \citep{comastri04, gilli04}.
While high column density AGN may be unidentifiable as AGN based on
their optical spectra alone, the fact that some fraction of their host
galaxies are optically bright (${\cal R}<25.5$) and fall within
optically-selected samples bodes well for determining their
spectroscopic redshifts.  Accurate spectroscopic redshifts are
particularly important for constraining AGN column densities ($N_{\rm
H}$); the inferred $N_{\rm H}$ depends strongly on the assumed
redshift, $N_{\rm H}(z) \approx N_{\rm H}(0) (1+z)^{2.6}$
\citep{alexander05}.

In summary, of the 11 AGN with $z>1.4$ in our optically-selected
sample, $7/11$ ($64\%$) are detected in X-rays to $2$~Ms, $9/11$
($82\%$) are detected with $8$ and/or $24$~$\mu$m excesses, and $8/11$
($73\%$) have rest-frame UV signatures of AGN.  Even in the deepest
X-ray image available, there is still a considerable number of AGN
that remain undetected, and we must incorporate other techniques,
e.g., optical spectra and $8$ and $24$~$\mu$m data, to fully account
for the census of AGN.

\section{Summary}
\label{sec:summary}

We have presented the results of a spectroscopic survey of redshift
$1.4\la z\la 3.0$ star-forming galaxies in the GOODS-North field, made
possible by efficient UV ($\ugr$) color selection and the unique
multi-object capabilities of the LRIS instrument on the Keck I
telescope.  Our sample consists of 212 redshifts for galaxies at
redshifts $1.4\la z\la 2.5$ selected using the BM and BX criteria of
\citet{adelberger04} and \citet{steidel04}, and 30 new redshifts (of a
total of 74) for Lyman break galaxies at redshifts $2.5\la z\la 3.5$.
Our deep optical and near-IR imaging, supplemented by publicly
available {\it Spitzer} IRAC and MIPS data (Dickinson et~al., in prep;
Chary et~al., in prep), allow us to measure the stellar populations,
stellar masses, star formation rates, and dust extinction for galaxies
in our sample (e.g., \citealt{erb06b, shapley03, shapley05, reddy04,
reddy05a, reddy06a, steidel04}).  These analyses indicate that the
$\ugr$-selected sample consists of galaxies which span two orders of
magnitude in age and stellar mass, and 4 orders of magnitude in dust
obscuration ($L_{\rm bol}/L_{\rm UV}$).  Included are galaxies with
bolometric star formation rates ranging from $\sim 5$~$\sfr$ to
$>1000$~$\sfr$.  We further identify at least 3 of 11 AGN in our
sample which appear to be heavily dust-obscured based on their
power-law SEDs longward of $2$~$\mu$m (rest-frame) and lack of
detection in the deep {\it Chandra} 2~Ms data \citep{alexander03}.  A
compilation of the multi-wavelength data for these 11 AGN indicates
that optical and {\it Spitzer} data are able to more efficiently (in
terms of integration time) select AGN at $z>1.4$ than X-ray data, but
optical spectra and {\it Spitzer} and {\it Chandra} data are all
required to fully account for the census of AGN at high redshifts.
The photometry and SED fitting results for galaxies in our sample are
available at http://www.astro.caltech.edu/$\sim$drlaw/GOODS/.

Large spectroscopic samples at high redshifts allow for a number of
other detailed investigations such as the galaxy and AGN/QSO
luminosity functions (\citealt{steidel99, adel00, shapley01, hunt04},
Reddy et~al. in prep.); metallicities \citep{pettini98, pettini01,
shapley04, erb06a}; signatures of galaxy feedback and IGM metal
enrichment \citep{adel03}; and accurate clustering analyses
\citep{adelberger05a, adelberger05b}.  This large range in galaxy
evolution studies highlights the versatility and efficiency of
optically-selected samples in addressing many fundamental issues in
cosmology.

\acknowledgements

We thank David Law for setting up the website where the galaxy
photometry and SED fits are available to the public.  We are grateful
to the staff of the Keck and Palomar Observatories for their help in
obtaining the data presented here.  This work has been supported by
grant AST 03-07263 from the National Science Foundation and by the
David and Lucile Packard Foundation.

\bibliographystyle{apj}
%\bibliography{apj-jour,myrefs}

\clearpage

\vskip 2in
\begin{deluxetable}{lccccccc}
\tabletypesize{\footnotesize}
\tablewidth{0pc}
\tablecaption{Sample Statistics to $\rs=25.5$}
\tablehead{
\colhead{} &
\colhead{} &
\colhead{$\rho_{\rm cand}$\tablenotemark{b}} &
\colhead{} &
\colhead{} &
\colhead{} &
\colhead{} &
\colhead{} \\
\colhead{Candidates} &
\colhead{$N_{\rm cand}$\tablenotemark{a}} &
\colhead{(arcmin$^{-2}$)} & 
\colhead{$N_{\rm obs}$\tablenotemark{c}} &
\colhead{$N_{\rm z>1}$\tablenotemark{d}} &
\colhead{$f_{\rm z>1}$\tablenotemark{e}} &
\colhead{$N_{\rm z>1.4}$\tablenotemark{f}} &
\colhead{$\langle z\rangle$\tablenotemark{g}}}
\startdata
BM & 470 & $3.13\pm0.14$ & 67 & 63 & 0.94 & 49 & $1.72\pm0.34$ \\
BX & 890 & $5.93\pm0.20$ & 205 & 170 & 0.83 & 163 & $2.19\pm0.40$ \\
C & 55 & $0.37\pm0.05$ & 26 & 26 & 0.97 & 26 & $3.11\pm0.21$ \\
D & 59 & $0.39\pm0.05$ & 23 & 23 & 0.98 & 23 & $3.05\pm0.22$ \\
MD & 78 & $0.52\pm0.06$ & 26 & 25 & 0.96 & 25 & $2.96\pm0.30$ \\
{\bf Total} & 1552 & $10.35\pm0.26$ & 347 & 307 & 0.88 & 286 & $2.25\pm0.57$ \\
\enddata
\tablenotetext{a}{Number of photometric candidates.}
\tablenotetext{b}{Surface density of photometric candidates to $\rs=25.5$.}
\tablenotetext{c}{Number of objects with secure spectroscopic identifications.}
\tablenotetext{d}{Number of objects with secure redshifts $z>1$.}
\tablenotetext{e}{Fraction of spectroscopically observed objects with
$z>1$.  The foreground ($z<1$) contamination rates of the C, D, and MD
samples are very low ($<5\%$) and we assume the interloper fractions
derived over all fields of the $z\sim 3$ Lyman Break Galaxy survey 
\citep{steidel03}.  For the BM and BX samples, we assume foreground 
contamination fractions
derived from the GOODS-N field, which are similar to those derived in other
fields of the $z\sim 2$ survey \citep{steidel04}.}
\tablenotetext{f}{Number of objects with $z>1.4$}
\tablenotetext{g}{Mean and standard deviation of redshift distribution
for objects with $z>1$.}
\label{tab:counts}
\end{deluxetable}
\clearpage
\LongTables
\begin{landscape}
\begin{deluxetable}{lcccccccccccccccl}
%\landscape
\tabletypesize{\tiny}
\tablewidth{0pc}
\tablecaption{GOODS-N $\ugr$ Galaxies with Spectroscopic Redshifts}
\tablehead{
\colhead{} &
\colhead{$\alpha$} &
\colhead{$\delta$} &
\colhead{} &
\colhead{} &
\colhead{} &
\colhead{${\cal R}$\tablenotemark{d}} &
\colhead{$\gmr$} &
\colhead{$\umg$\tablenotemark{e}} &
\colhead{$J$\tablenotemark{f}} &
\colhead{$\ks$\tablenotemark{f}} &
\colhead{m$_{\rm 3.6\mu m}$} &
\colhead{m$_{\rm 4.5\mu m}$} &
\colhead{m$_{\rm 5.8\mu m}$} &
\colhead{m$_{\rm 8.0\mu m}$} &
\colhead{f$_{\rm 24\mu m}$} & 
\colhead{} \\
\colhead{Name} &
\colhead{(J2000.0)} &
\colhead{(J2000.0)} &
\colhead{$z_{\rm em}$\tablenotemark{a}} &
\colhead{$z_{\rm abs}$\tablenotemark{b}} &
\colhead{Type\tablenotemark{c}} &
\colhead{(mag)} &
\colhead{(mag)} &
\colhead{(mag)} &
\colhead{(mag)} &
\colhead{(mag)} &
\colhead{(mag)} &
\colhead{(mag)} &
\colhead{(mag)} &
\colhead{(mag)} &
\colhead{($\mu$Jy)} &
\colhead{Notes\tablenotemark{g}}}
\startdata
BX1035 & 12:36:13.03 & 62:10:21.1 & ... & 2.236 & GAL & 23.46 & 0.37 & 0.68 & ... & ... & 22.48 & 22.47 & 22.11 & 22.59 & 33.0 	\\     
BX1040 & 12:36:17.81 & 62:10:11.2 & 2.469 & 2.466 & GAL & 24.84 & 0.22 & 0.68 & ... & ... & 22.94 & 23.27 & ... & ... & ... 	\\     
BX1042 & 12:35:50.89 & 62:13:33.5 & 2.613 & 2.601 & GAL & 24.83 & 0.50 & 1.16 & ... & ... & 23.09 & 23.03 & 23.13 & 22.88 & 15.2 	\\     
BX1050 & 12:36:18.80 & 62:10:37.4 & ... & 2.322 & GAL & 24.71 & 0.40 & 0.73 & ... & ... & 23.49 & 23.51 & 23.46 & ... & ... 	\\     
BX1051 & 12:35:52.97 & 62:13:36.8 & ... & 2.098 & GAL & 24.23 & 0.14 & 0.57 & ... & ... & 23.07 & ... & ... & ... & 203.9 	\\     
BX1055 & 12:35:59.59 & 62:13:07.5 & 2.496 & 2.486 & GAL & 24.09 & 0.24 & 0.81 & ... & ... & 23.12 & 23.23 & ... & ... & ... 	\\     
BX1060 & 12:36:06.40 & 62:12:29.1 & ... & 2.081 & GAL & 24.22 & 0.43 & 0.84 & ... & ... & 21.74 & 21.73 & 21.93 & ... & 67.3 	\\     
BX1064 & 12:36:30.34 & 62:09:45.3 & ... & 2.086 & GAL & 24.20 & 0.28 & 0.61 & 23.60 & 22.63 & 23.00 & 23.11 & 23.16 & 23.76 & 26.0 	\\
\enddata
\tablenotetext{*}{Note.---A full version of
this table, including magnitude and flux errors, is available in the accompanying machine-readable
table and will be available in the electronic edition of the
{\it Astrophysical Journal}.}
\tablenotetext{a}{Emission line redshift.  An entry with a colon indicates the redshift
is uncertain.  A blank entry indicates that an emission line redshift could not be measured.}
\tablenotetext{b}{Absorption line redshift.  An entry with a colon indicates the redshift
is uncertain.  A blank entry indicates that an absorption line redshift could not be measured.}
\tablenotetext{c}{Source type, either galaxy (``GAL''), ``AGN'', ``QSO'', or ``STAR''.  The distinction between QSO and AGN classification is based on the
line widths, as described in \citet{steidel02}, where an object is classified
as a ``QSO'' if it has any emission line with FWHM$>2000$~km~s$^{-1}$.  AGN
identified using other means (X-ray or mid-infrared emission) are discussed
in \S~\ref{sec:agn}.}
\tablenotetext{d}{${\cal R}$ magnitude in AB units.}
\tablenotetext{e}{Upper limits given for galaxies undetected in $U_{\rm n}$.}
\tablenotetext{f}{A blank entry indicates the object did not lie in the region with near-IR imaging.}
\tablenotetext{g}{Galaxies in common with the LBG survey in the HDF-N
  are indicated by their names (``S03-XXX'') as given in
  \citet{steidel03}.  Galaxies with X-ray counterparts within
  $1.5\arcsec$ are indicated by their names (``B03-XXX'') in the
  spectroscopic followup to the {\it Chandra} 2~Ms survey by
  \citet{barger03}.  Galaxies satisfying the distant red galaxy (DRG)
  criteria of \citet{franx03} are indicated by ``DRG''.}
\label{tab:galaxies}
\end{deluxetable}
\clearpage
\end{landscape}
\begin{deluxetable}{lccccccccccc}
\tabletypesize{\footnotesize}
\tablewidth{0pc}
\tablecaption{Stellar Population Parameters}
\tablehead{
\colhead{} &
\colhead{} &
\colhead{} &
\colhead{Age$_{\infty}$} &
\colhead{SFR$_{\infty}$} &
\colhead{M$^{\ast}_{\infty}$} &
\colhead{$\tau$\tablenotemark{b}} &
\colhead{} &
\colhead{Age$_{\tau}$} &
\colhead{SFR$_{\tau}$} &
\colhead{M$^{\ast}_{\tau}$} &
\colhead{}\\
\colhead{Name\tablenotemark{a}} &
\colhead{$z$} &
\colhead{$\ebmv_{\infty}$} &
\colhead{(Myr)} &
\colhead{($\sfr$)} &
\colhead{($10^{10}$~M$_{\odot}$)} &
\colhead{(Myr)} &
\colhead{$\ebmv_{\tau}$} &
\colhead{(Myr)} &
\colhead{($\sfr$)} &
\colhead{($10^{10}$~M$_{\odot}$)} &
\colhead{Notes}\tablenotemark{c}}
\startdata
BX1035 & 2.236 & 0.24 & 20 & 234 & 0.466 & 20 & 0.24 & 15 & 185 & 0.419 \\
BX1040 & 2.468 & 0.09 & 2100 & 14 & 3.030 & 200 & 0.03 & 571 & 7 & 2.220 \\
BX1042 & 2.607 & 0.18 & 509 & 36 & 1.820 & 200 & 0.15 & 321 & 23 & 1.810 \\
BX1050 & 2.322 & 0.27 & 15 & 118 & 0.179 & 50 & 0.26 & 15 & 98 & 0.173 \\
BX1051 & 2.098 & 0.12 & 404 & 24 & 0.963 & 100 & 0.02 & 286 & 7 & 1.100 \\
BX1055 & 2.491 & 0.12 & 255 & 37 & 0.956 & 100 & 0.04 & 227 & 14 & 1.210 \\
BX1060 & 2.081 & 0.24 & 806 & 64 & 5.150 & 100 & 0.01 & 509 & 4 & 5.810 \\
BX1064 & 2.086 & 0.14 & 509 & 22 & 1.100 & 100 & 0.02 & 321 & 5 & 1.230 & \\
\enddata
\tablenotetext{a}{We did not fit the stellar populations of galaxies
that had no data longward of ${\cal R}$-band, had uncertain redshifts,
or are identified as AGN/QSO from their optical spectra.  We also do not present
SED parameters for those galaxies with optical and IRAC photometry inconsistent
with a simple stellar population (these sources had large $\chi^2 > 10$ or those
sources with $8$~$\mu$m or $24$~$\mu$m excesses.}
\tablenotetext{b}{Best-fit star formation history decay timescale in Myr.  In
some cases, a model with constant star formation ($\tau=\infty$) provided 
the best-fit.} 
\tablenotetext{c}{Objects with large $\ks$-band residuals are
indicated (i.e., those sources whose $\ks$-band measurement lies more
than $3$~$\sigma$ away from the best-fit stellar population).  MD31 is
the only directly-detected X-ray source which shows no optical
signatures of an AGN, has a relatively faint $24$~$\mu$m flux, and has
photometry that is consistent with a stellar population.  The IRAC
photometry for BX1267 may suffer from a deblending problem and these
(IRAC) data were not used in the SED fit. }
\label{tab:sedparms}
\end{deluxetable}
\clearpage
\vskip 2in
\begin{deluxetable}{lccc}
\tabletypesize{\footnotesize}
\tablewidth{0pc}
\tablecaption{Stellar Masses}
\tablehead{
\colhead{Sample} &
\colhead{$N_{\rm sed}$\tablenotemark{a}} &
\colhead{$\log [M_{\rm med}^{\ast}/{\rm M_{\odot}}]$\tablenotemark{b}} &
\colhead{$\log [M^{\ast}/{\rm M_{\odot}}]$\tablenotemark{c}}}
\startdata
BM & 51 (3) & 9.87 (10.70) & $9.85\pm0.58$ ($10.58\pm0.51$) \\
BX & 157 (7) & 10.20 (10.98) & $10.15\pm0.52$ ($10.94\pm0.38$) \\
LBG & 46 (0) & 10.05 (...) & $10.03\pm0.44$ (...) \\
{\bf Total} & 254 (10) & 10.10 (10.98) & $10.06\pm0.53$ ($10.83\pm0.43$) \\
\enddata
\tablenotetext{a}{Number of galaxies for which we derived best-fit SED
parameters (including stellar mass), regardless of whether the galaxies
were imaged at $\ks$.  The parentheses indicate the number of galaxies with
{\it measured} $\ks<21.82$, or $\ks({\rm Vega})<20$.}
\tablenotetext{b}{Median stellar mass assuming a constant star formation
(CSF) history.  Numbers in parentheses give the median stellar mass for
those galaxies with $\ks<21.82$, or $\ks({\rm Vega})<20$.}
\tablenotetext{c}{Mean and dispersion of stellar mass distribution assuming
a CSF history.  Numbers in parentheses give the mean stellar mass and dispersion
for those galaxies with $\ks<21.82$, or $\ks({\rm Vega})<20$.}
\label{tab:masses}
\end{deluxetable}

\vskip 1in
\begin{deluxetable}{lcccccc}
\tabletypesize{\footnotesize}
\tablewidth{0pc}
\tablecaption{AGN at $z>1.4$}
\tablehead{
\colhead{Name} &
\colhead{z} & 
\colhead{X-ray?\tablenotemark{a}} &
\colhead{Optical?\tablenotemark{b}} &
\colhead{$8$~$\mu$m?\tablenotemark{c}} &
\colhead{$24$~$\mu$m?\tablenotemark{d}} & 
\colhead{Notes\tablenotemark{e}}}
\startdata
BX1637 & 2.487 & no & yes (AGN) & yes & yes & ACS point source \\
BX160 & 2.460 & no & yes (AGN) & yes & yes & $\ks=21.87$ \\
BM1083 & 2.414 & yes & yes (QSO) & yes & yes & \\
BM1119 & 1.717 & yes & no & yes & yes & $\ks=21.74$ \\
BM1156 & 2.211 & no\tablenotemark{f} & yes (AGN) & yes & yes & very weak X-ray source \\
BM1384 & 2.243 & yes & yes (AGN) & yes & no & $\ks=21.69$ \\
MD31 & 2.981 & yes & no & no & no & ACS point source \\
MD39 & 2.583 & yes & yes (QSO) & yes & yes & $\ks=20.07$ \\
MD74 & 2.635 & no & no & yes & yes \\
S03-oC34 & 3.413 & yes & yes (QS0) & no & no \\
S03-MD12 & 2.647 & yes & yes (AGN) & yes & yes \\
\enddata
\tablenotetext{a}{Indicates if the source lies within $1\farcs 5$ of
an X-ray counterpart in the $2$~Ms data \citep{alexander03}.}
\tablenotetext{b}{Indicates if the source shows high ionization 
optical emission lines indicative of AGN/QSO.}
\tablenotetext{c}{Indicates if the source has a significant $8$~$\mu$m excess
compared with the flux at shorter wavelengths, larger than would be expected 
from a single stellar population.}
\tablenotetext{d}{Indicates if the source has a $24$~$\mu$m flux
$f_{\rm 24\mu m}>100$~$\mu$Jy.}
\tablenotetext{e}{$\ks$ magnitudes are indicated for AGN with $\ks<21.82$,
or $\ks({\rm Vega})\la 20$.}
\tablenotetext{f}{BM1156 is not included in the {\it Chandra} 2~Ms catalogs,
but further refinements to the X-ray data reduction procedure yielded a very
weak detection (N. Brandt, private communication).}
\label{tab:agn}
\end{deluxetable}

\end{document}